\documentclass{jnmp}

\usepackage{amsmath}
\JNMPnumberwithin{equation}{section}

\newtheorem{theorem}{Theorem}
\newtheorem{lemma}{Lemma}

\theoremstyle{definition}
\newtheorem{definition}{Definition}
\newtheorem{remark}{Remark}

\begin{document}

\renewcommand{\evenhead}{A Bellouquid}
\renewcommand{\oddhead}{The Incompressible Navier--Stokes
for the Non Linear Discrete Velocity Models}

\thispagestyle{empty}

\FirstPageHead{9}{4}{2002}{\pageref{Bellouquid-firstpage}--\pageref{Bellouquid-lastpage}}{Article}

\copyrightnote{2002}{A Bellouquid}

\Name{The Incompressible Navier--Stokes \\
for the Nonlinear Discrete Velocity Models}
\label{Bellouquid-firstpage}

\Author{A BELLOUQUID}

\Address{Department of Mathematics, Politecnico of Torino,\\
Corso   Duca degli Abruzzi 24, 10129, Torino, Italy\\
E-mail: bellouq@calvino.polito.it}

\Date{Received September 30, 2001; Revised March 18, 2002;
Accepted April 25, 2002}

\begin{abstract}
\noindent
We establish the incompressible  Navier--Stokes limit
  for the   discrete velocity model of the Boltzmann equation
  in any dimension of the physical space,
     for densities which remain in a suitable
small neighborhood of the global Maxwellian.
 Appropriately scaled families solutions of discrete
Boltzmann equation
are shown to have fluctuations that locally in time converge  strongly
to a limit governed by a solution of  Incompressible
Navier--Stokes provided that the initial fluctuation is smooth, and
  converges
   to appropriate initial data.
As applications of
our results, we study the Carleman model and
the one-dimensional Broadwell model.
\end{abstract}

\section{Introduction and Main results}

\subsection{Introduction}

The endeavor to understand how fluid dynamical equations can be derived from
  kinetic theory goes back to the founding works of Maxwell [38] and Boltzmann [16].
Most of derivations   are well understood at several formal levels by now,
and yet their full mathematical justifications are still missing.
  Here we establish a so-called incompressible Navier--Stokes
fluid dynamical limit for the discrete
  Boltzmann equation  which is
  an evolution model
  for a gas which can attain only a finite number of velocities, one has to mention that  Inoue
  and Nishida~[30],
  and Caflish and  Papanicolaou~[19] have developed an asymptotic
theory for the six velocity Broadwell's model. Further Studies have
 been developed for the two velocity
 Carleman's model by Kurtz~[33]  and by  Lions and Toscani~[36].
The interested reader can recover the pertinent literature in the survey by Lachowicz~[34].

As known, see Bonilla and Soler~[14], different asymptotic expansions can be developed for
kinetic models  thus obtaining various hydrodynamic descriptions. It has  been
 recently  recognized by several mathematicians, Bardos,
 Golse and Levermore~[3,~4],
  Demasi,   Esposito and Lebowitz~[21], that the incompressible Navier--Stokes equation can be
 obtained as the limit of the Boltzmann equation, when
 the Knudsen number $\varepsilon >0 $
  go to zero.  The
  validity of this fluid-dynamical approximation
    have been studied
   by several authors, see~[2, 5, 6, 10, 21, 25, 26, 37]. On the other hand, there is non
proof concerning the  convergence of
 discrete velocity kinetic models to
 a solution of the
incompressible Navier--Stokes equations. However, although a complete proof is not given,
some partial results may be found in~[9, 10, 11, 28]
 in particular for
Carleman and Broadwell's models.
 In the present paper a rigorous proof of the connection between
these two points of view is done, precising and
completing some results  which were announced in~[4].

Let $\varepsilon >0$ be the   Knudsen number,
 and consider the scaled discrete
 Boltzmann equation
\begin{equation}
\varepsilon \partial_t F_i + v_i \cdot \nabla_x F_i
    =  \frac{1}{\varepsilon} Q_i(F,F),\qquad i=1,\ldots,m,
\end{equation}
where $F_i =F_i(t,x)$ represents the mass density of gas particles
with the constant velocity vector $v_i
=(v_i^1,\ldots,v_i^n)\in{\mathbb R}^n$  at time $t\geq 0$ and
position $x=(x_1,\ldots, x_n)\in{\mathbb R}^n$; $Q_i$ is
a~quadratic operator related to the binary collisions:
\begin{equation}
Q_i (F,G)= \frac{1}{2\alpha_i} \sum_{jkl} \left\{A_{kl}^{ij} (F_k G_l +
F_lG_k)-A_{ij}^{kl}(F_iG_j +F_jG_i)\right\},
\end{equation}
where $\alpha_i$ are positive constants,
and the terms $A_{ij}^{kl}$ are the so-called  transition rates referring to the collisions
\[
(v_i, v_j) \leftrightarrow   (v_k,v_l).
\]
The transition rates are positive constants which, according to
the indistinguishability property of the gas particles
and the reversibility of the collisions, fulfill the following properties:
\begin{equation}
A_{lk}^{ij}=A_{kl}^{ij}=A_{kl}^{ji}=\cdots =A_{ij}^{kl}.
\end{equation}

A detailed computation of the terms
$A_{ij}^{kl}$ can be performed by specializing the velocity discretization and
analyzing the related collision mechanics.

The above mathematical structure includes,
as particular applications, various mathematical models
of discrete
 Boltzmann equation which can be recovered in the pertinent literature [18, 23].
 It also refers to models
 with arbitrarily large number of velocities which have been studied in a number
 of recent papers, see among others [1, 27], considering their
  relatively simpler structure, with respect
 to the full Boltzmann equation, which may
be exploited for scientific computing. Indeed, convergence
to the full Boltzmann equation can be studied for models with
 arbitrary  large number of velocities.

Following [7, 18, 23],
 we shall introduce the basic concepts concerning (1.1) and
summarize their properties which will be used later.

\begin{definition}
   A vector $\phi =t_{(\phi_1,\ldots,\phi_m)}\in {\mathbb R}^m$
is called a summational invariant if
\[
A_{kl}^{ij}\left(\frac{\phi_i}{\alpha_i}+\frac{\phi_j}{\alpha_j}-\frac{\phi_k}{\alpha_k}
-\frac{\phi_l}{\alpha_l}\right)=0,
\]
for all $i,j,k,l=1,\ldots,m$.
\end{definition}

We denote by $\mathbb M$ the set of summational invariants. Then
$0<\dim {\mathbb M}<m$ because $t_{(\alpha_1,\ldots,\alpha_m)}\in
{\mathbb M}$ and ${\mathbb M}\neq {\mathbb R}^m$.
 The
following three conditions are equivalent:

\begin{enumerate}
\vspace{-2mm}
\itemsep0mm
\item[(i)] $\phi \in {\mathbb M}$,

\item[(ii)] $\langle \phi, Q(F,G)\rangle =0$ for all $F,G \in {\mathbb R}^m$,

\item[(iii)] $\langle \phi, Q(F,F)\rangle =0$ for all $F \in {\mathbb R}^m$.
\vspace{-2mm}
\end{enumerate}

Here $\langle \cdot, \cdot\rangle$ denotes the standard inner product in ${\mathbb R}^m$.
For the proof see   [7, 18, 23].

Let $F=t_{(F_1,\ldots,F_m)}\in {\mathbb R}^m$. We write $F>0$ if $F_i >0$ for all $i=1,\ldots,m$.

\begin{definition} A vector $F=t_{(F_1,\ldots,F_m)}>0$ is called a local Maxwellian if
\[
A_{kl}^{ij}(F_i F_j -F_kF_l)=0,  \qquad \mbox{for all}\quad i,j,k,l=1,\ldots,m.
\]
In particular, $F>0$ is called an absolute Maxwellian if it is a locally
Maxwellian state and is independent of $t$ and $x$.
\end{definition}

 Consider the initial value problem for Eq. (1.1):
\begin{gather}
\varepsilon \partial_t F_\varepsilon +
\sum_{j=1}^{n}V^j\partial_{x_j}F_\varepsilon =
\frac{1}{\varepsilon} Q(F_\varepsilon,F_\varepsilon), \qquad t>0,
\quad x\in {\mathbb R}^n,\nonumber\\
 F_\varepsilon(0,x)= F_0(x),
\end{gather}
where
\[
V^j = {\rm diag}\left(v_j^1,\ldots,v_j^m\right), \qquad j=1,\ldots,n,
\]
and
\[
Q(F,G)=t_{(Q_1(F,G),\ldots,Q_m(F,G))}.
\]

According to  [4], the discrete velocity Boltzmann equation (1.1)
gives  the incompressible Navier--Stokes equation in the limit
$\varepsilon \to 0$ if $F_\varepsilon$ remains near an absolute
Maxwellian $M$ with the distance of order~$\varepsilon$.

Let $M>0$ be an absolute Maxwellian state, and
let
\[
\Lambda = {\rm diag} \left(\frac{M_1}{\alpha_1},\ldots,\frac{M_m}{\alpha_m} \right).
\]
 We shall see the solution in the form
\[
F_\varepsilon (t,x)=M
+\varepsilon\Lambda^{\frac{1}{2}}f_\varepsilon.
\]
Then Problem  (1.4)  can be transformed into the following one:
\begin{gather}
\partial_t f_\varepsilon + \frac{1}{\varepsilon}\sum_{j=1}^{n}
V^j\partial_{x_j}f_\varepsilon
+\frac{1}{\varepsilon^2}Lf_\varepsilon= \frac{1}
{\varepsilon}\Gamma (f_\varepsilon,f_\varepsilon), \nonumber\\
f_\varepsilon(t=0,x)=f_0(x) =\Lambda^{-\frac{1}{2}}(F_0(x)-M),
\end{gather}
where  the operators $L$ and $\Gamma$ are given by
\begin{gather}
Lf=-2\Lambda^{-\frac{1}{2}}Q\left(M, \Lambda^{\frac{1}{2}}f\right),\\
\Gamma (f,g)= \Lambda^{-\frac{1}{2}}Q\left(\Lambda^{\frac{1}{2}}f,\Lambda^{\frac{1}{2}}g\right),
\end{gather}
and  have (see [7, 18, 23])
the following properties:
\begin{enumerate}
\vspace{-2mm}
\itemsep0mm
\item[(i)] $L$ is real symmetric and positive semi-definite.
It is null space is given by
$N(L)= \Lambda^{\frac{1}{2}}{\mathbb M}$.  We denote by $\{e_i, i=1,\ldots,d \}$
the orthonormal basis for $N(L)$.
\item[(ii)] $\Gamma$ is bi-linear and satisfies $\Gamma(f,g) \in N(L)^\perp$ for any $f,g \in
 {\mathbb R}^m$, where
$N(L)^\perp$  denotes the orthogonal complement of $N(L)$ in ${\mathbb R}^m$.
\vspace{-2mm}
\end{enumerate}

Existence of global solutions for discrete kinetic models have
been obtained by several authors (see [15, 17] and the review
article of Bellomo and Gustafsson~[8]). However these results
cannot be used here because they are based on the dispersion
properties of the linearized equations and concern small (with
respect to the collision operator) perturbations of the vacuum. It
is easy to prove that any finite velocity model has a~unique
smooth solution during a finite time that depends on the size of
the initial data and on the collision operator (or a global
existence see [31, 32]). To the best of our knowledge it has not
been proven that the time of existence of this smooth solution is
independent of~$\varepsilon$.

Now we  state our main results.

\subsection{Main results}

Our main result is an existence theorem that holds for all
$\varepsilon >0 $  and a proof of the validity of the
fluid-dynamical approximation. To state our result precisely, one
 needs
 some function spaces.

 Let $C(\Omega, X) $
 and $L^\infty (\Omega,X)$ denote the spaces of
 the  continuous and bounded functions on $\Omega\subset{\mathbb R}^n$ with values in a
Banach space $X$, respectively.

 $H^l$ denotes the $L^2\left({\mathbb R}^n\right)$-Sobolev space of order $l$,
with the norm  $\|\cdot\|_{l}$.

\begin{theorem} Let (1.3) be assumed. Let $n\geq 1$
and $l\geq \frac{n}{2}$. If $f_0\in H^l\left({\mathbb R}^n\right)$,
then there exists a positive constants $T_0$ and $k$
 (depending only on $\| f_0\|_l$)
such that the initial value problem (1.5) has a unique solution
$f_\varepsilon \in L^{\infty} \left([0,T_0], H^l\right) \cap
C\left([0,T_0], H^{l-1}\left( {\mathbb R}^n\right)\right)$
satisfying
\begin{equation}
 \| f_\varepsilon (t) \|_l \leq k,
\end{equation}
 for $t\in [0,T_0]$.
\end{theorem}

 If, in addition  the initial data satisfies:
\begin{gather}
f_\varepsilon (0)= h_\varepsilon +\varepsilon^2 k_\varepsilon
\qquad \mbox{where}
 \quad k_\varepsilon \in H^l \qquad \mbox{and,}\nonumber\\
 h_\varepsilon\in N(L), \qquad \| \Gamma(h_\varepsilon,h_\varepsilon)\|_{l-1}\leq C\varepsilon,
\qquad \| \partial_x h_\varepsilon   \|_{l-1} \leq C
\varepsilon,\nonumber\\
 \lim_{\varepsilon \rightarrow 0}\|
h_\varepsilon -h \|_{l-1} = 0.
\end{gather}
one gets the strong convergence for the discrete
Boltzmann equation to Incompressible Navier--Stokes equations:

\begin{theorem}  Let $f_\varepsilon $ be as in Theorem 1. Then, as $\varepsilon\rightarrow 0$,
 $f_\varepsilon\rightarrow f$  weakly  $\star$   in  \linebreak  $L^\infty \left([0,T], H^{l} \right)$
 and  strongly in
$C\left( [0, T], H^{l-1}\right)$ for  any $T>0$, and the
  limit has the form
\begin{equation}
f = \sum_{i=1}^{d} \rho_i e_i,
\end{equation}
where $\rho_i = \langle e_i,f\rangle$ satisfy:
\begin{equation}
\nabla_x \cdot \langle V e\otimes e \cdot \rho\rangle =0.
\end{equation}
Moreover, $\rho (t,x)$ is a weak solution of the equations:
\begin{gather}
\frac{\partial\rho}{\partial t} - \nabla^2 : (\langle
L^{-1}(P^\perp (Ve))\otimes (Ve)\rangle \cdot \rho )+ \frac{1}{2}
\nabla\cdot \langle P^\perp (Ve)\frac{(\rho\cdot
e)^2}{\sqrt{\alpha M}}\rangle =
 \langle V\cdot \nabla e \otimes e \rangle \cdot \pi,\nonumber\\
\rho (t=0)= \langle h,  e \rangle,
\end{gather}
where $\frac{g^2}{\sqrt{\alpha M}}$ denotes the vector
$\left(\frac{g_1^2}{\sqrt{\alpha_1 M_1}},\ldots,
 \frac{g_n^2}{\sqrt{\alpha_n M_n}}\right)$ and the
 vectors
\[
 \nabla^2 : \langle
 Ve \otimes L^{-1}(P^\perp(Ve))\rangle \rho\qquad \mbox{and}\qquad
\nabla\cdot \langle P^\perp (Ve),
\frac{(\rho\cdot e)^2}{\sqrt{\alpha M}}\rangle
\]
 are given by:
\begin{gather*}
\nabla^2 : \langle Ve \otimes L^{-1}(P^\perp(Ve))\rangle \rho =
 \left(\sum_{j=1}^{n}\sum_{k=1}^{n}\partial_{x_j} \partial_{x_k}\langle
V^k  \rho\cdot e ,L^{-1}\left(P^\perp
\left(V^je_i\right)\right)\rangle\right)_{i},\\ \nabla\cdot
\langle P^\perp (Ve),\frac{(\rho\cdot e)^2}{\sqrt{\alpha
M}}\rangle= \left(\sum_{j=1}^{n}\partial_{x_j}\langle
\frac{(\rho\cdot e)^2} {\sqrt{\alpha M}},P^\perp
\left(V^je_i\right)\rangle\right)_{i}.
\end{gather*}
\end{theorem}

Let $ (\rho_\varepsilon)_i=  \langle e_i,f_\varepsilon\rangle$.
Since $\{e_i, i=1,\ldots,d\}$ forms an orthogonal system, $\rho$
in (1.10) is given by   $\rho_i=  \langle e_i,f\rangle$. One gets

\begin{theorem}
  Let (1.9) be assumed.
  Then, as $\varepsilon\to 0$,
 $\rho_\varepsilon\to \rho  $  weakly  $\star$   in $L^\infty \left([0,T], H^{l} \right)$
 and  strongly  in
$C\left( [0, T], H^{l-1}\right)$ for  any  $T>0$,  and the limit $\rho$
satisfies the incompressible Navier--Stokes equations (1.11)--(1.12).
\end{theorem}

 The system (1.11)--(1.12) are generalizations of the
 Navier--Stokes equations.
 The condition (1.11) generalizes the $\nabla\cdot u=0$, $\nabla(\rho+\theta)=0$
 conditions that arise in the classical incompressible
 Navier--Stokes limit (see [3]). $\pi$ is the lagrange
 multiplier corresponding to the constraint (1.11)
 and is the generalization of the classical pressure term.    The quadratic
 term in (1.12) corresponds to the classical convection
 terms.
 The justification of  these
  formal approximations for the classical Boltzmann equation
   has proven difficult
because many regularity questions remain open for both these fluid systems
as the Boltzmann equation.
 Two approaches
 to circumventing these difficulties have emerged recently. First
 some
  authors have studied direct derivations of linear or weakly nonlinear fluid
dynamical systems, such as
 incompressible Navier--Stokes~[21].  Their result requires smooth initial data and holds
 for as long as the limiting
  solution of the  incompressible Navier--Stokes system is smooth.
Second, some authors have abandoned
the traditional expansion-based derivations in favor of
 moments based formal derivation~[5, 6, 26].
  In~[6] it is shown that the solution
 of the Boltzmann equation considered over space
 of dimension three
  or more will be smooth for all time,
  with  small
   initial
 data and will converge   strongly
  to the solution of the  incompressible Navier--Stokes equations. Recently
 the  Incompressible Navier--Stokes fluid dynamical limit for the classical
 Boltzmann equation is considered  in~[26].  It was shown  that the
 scaled families of
 DiPerna--Lions renormalized
 solutions~[22]
 have fluctuations that globally in time converge weakly to a  limit
 governed by a solution of  the incompressible Navier--Stokes
  equations due  to Leray~[35]
   provided that
\[ H (f_\varepsilon (0)) = \iint_{{\mathbb R}^3\times{\mathbb R}^3 }
    (f_\varepsilon(0)\log f_\varepsilon(0) -
   f_\varepsilon(0)+1 )M\, dvdx \leq C\varepsilon^2,
\]
 and
  the fluid moments of their initial fluctuations
 converge to appropriate  $L^2$ initial data.
 The proof use
  the averaging lemma (cf.~[24]). The averaging lemma is valid for
 continuous solutions and has no counterpart for discrete velocity models (except in one
 space dimension (cf.~Tartar~[39])). However, in the proof of Theorem~2,
one needs to evaluate the limit of
 the nonlinear moment
\begin{equation}
\langle L^{-1}(P^\perp (Ve)),
\Gamma(f_\varepsilon,f_\varepsilon)\rangle.
\end{equation}
  Some uniform regularity estimates would likely be needed
 for obtaining the limit of nonlinear terms (1.13).  This term disappears
 in the case of Stokes limit (see~[12]).

 Here we establish a so-called incompressible Navier--Stokes fluid dynamical
limit  (1.11)--(1.12)
 for the discrete Boltzmann
equation in any dimension    of the physical space.  We solve~(1.5) by using
the principle of contraction mappings,
    we introduce the iteration scheme  (Lemma~1 and Lemma~2)
    similar to one used in~[19] to obtain the uniform estimate of
     the remainder
    term in the   Hilbert  expansion in order
    to justify the compressible Euler system from the
    one dimensionnal Broadwell.
The convergence of the scheme to a solution of equation (1.5) is proved
    by using some properties of the operators $L$ and $\Gamma$.
    The strong convergence of the solution of equation
(1.5) as  $\varepsilon \to 0$ is proved by the uniform estimate
and the equicontinuity in $t\in [0,T]$ of the solution with
respect to $\varepsilon \in (0,1)$  (Lemma~3)
 provided
that the initial fluctuation is smmoth,
    close
   to an $N(L)$ element  which converges
    to appropriate initial data.

\begin{remark}
Put $h_\varepsilon = \Lambda^{\frac{1}{2}} t_{(\alpha_1,
\alpha_2,\ldots,\alpha_m)}w_\varepsilon$, one gets
\begin{gather}
 \Gamma (h_\varepsilon,h_\varepsilon) = w_\varepsilon^2 \Lambda^{-\frac{1}{2}}
Q(M, M) =0.
\end{gather}
An
 example of assumptions (1.9)    can be
 given by:
\begin{gather}
f_\varepsilon = \Lambda^{\frac{1}{2}}t_{(\alpha_1,
\alpha_2,\ldots,\alpha_m)} w_\varepsilon+\varepsilon^2
k,\nonumber\\
  \| \partial_x w_\varepsilon   \|_{l-1} \leq C \varepsilon,
  \qquad \lim_{\varepsilon \rightarrow 0}\| w_\varepsilon -w\|_{l-1} = 0.
\end{gather}
\end{remark}

\begin{remark}
 In the case where  $\frac{h_\varepsilon^2}{\sqrt{\alpha M}} \in N(L)$, one has
   $\Gamma(h_\varepsilon,h_\varepsilon) = 0 $ (see Lemma~5).
Therefore an other example can be given by:
\begin{gather*}
f_\varepsilon (0)= h_\varepsilon +\varepsilon^2
k_\varepsilon,\qquad \mbox{where} \quad k_\varepsilon \in H^l
\qquad \mbox{and},\\
 h_\varepsilon\in N(L), \qquad \frac{h_\varepsilon^2}{\sqrt{\alpha M}} \in N(L) ,
\qquad    \| \partial_x h_\varepsilon  \|_{l-1} \leq C
\varepsilon, \qquad \lim_{\varepsilon \rightarrow 0}\|
h_\varepsilon -h \|_{l-1} = 0.
\end{gather*}
\end{remark}

\begin{remark}
 In [12] it was proven that the kinetic models (1.1) converge weakly and
  strongly  to the linearized
 incompressible Navier--Stokes.
 The assumption $\Gamma(h_\varepsilon,h_\varepsilon) = O(\varepsilon)$ in $H^{l-1}$
  is not necessary in this case
  and the time of existence of solutions
  goes to infinity as $\varepsilon \to 0$.
\end{remark}

\begin{remark}
  The present work  improves upon
  the result given in [12] for  the
 Carleman and
    Broadwell's
    model, without
   assuming some restriction  upon the scaling of the fluctuation with
   respect to Knudsen number, namely that the fluctuations should only be required to be of
   an order equal  to
   the Knudsen number, but at the cost of being  restricted to initial
   fluctuations (assuming $\Gamma(h_\varepsilon,h_\varepsilon) = O(\varepsilon)$ in $H^{l-1}$)
   for the Broadwell model.
\end{remark}

 The plan of this paper is as follows. Section~1
deals with the above introduction and main results.
In Section~2, the formulation of the problem
and the proof of uniform
 existence Theorem are given. In Section~3, the
estimate for $\partial_t f_\varepsilon$ is shown. This estimate is
used to prove the strong convergence of the solution
 to the solution of
the non-linear
incompressible Navier--Stokes
equation. Finally, Sections~4 and~5 contains some applications.
As applications of our results, we are
 dealt   with the Carleman model and
the one-dimensional Broadwell model.

\section{Uniform existence}

To prove the existence of local solutions to (1.5), one has to
 get  suitable ``apriori'' estimate.

\subsection{Estimates}

\begin{lemma}Let  $z(t,x)$ be a given function of $t$ and $x$ such that,
\begin{equation}
\| z(t)\|_l \leq k,
\end{equation}
and let $f(t,x)$ satisfy the linear system
\begin{gather}
\partial_t f + \frac{1}{\varepsilon}\sum_{j=1}^{n}
V^j\partial_{x_j}f +\frac{1}{\varepsilon^2}Lf= \frac{1}
{\varepsilon}\Gamma (z,f),\nonumber\\
 f(0,x)=f_0(x).
\end{gather}
Then there exist $T_0$ such that   a constant
 $k$ can be chosen such
 that
\begin{equation}
\sup\limits_{0\leq t\leq T_0}\|f(t)\|_l \leq k.
\end{equation}
\end{lemma}

\begin{proof}
From the theory of linear hyperbolic systems  we
 know that
 (2.2) has a unique solution in $L^\infty \left([0,T], H^l\right)$ with
 $\frac{df}{dt}$ in $L^\infty \left([0,T],L^2\right)$.
X-Mozilla-Status: 0000

 Taking the Fourier  transform of (2.2) in $x$ yields
\begin{equation}
\partial_t \hat f + \frac{1}{\varepsilon}\sum_{j=1}^{n}
V^j i\zeta_j \hat f +\frac{1}{\varepsilon^2}L\hat f= \frac{1}
{\varepsilon}\hat\Gamma (z,f).
\end{equation}
Take the inner product (in ${\mathbb C}^m$ ) of (2.4) with $\hat f$. Since $\sum\limits_{j=1}^{n}
V^j \zeta_j$ and $L$ are real symmetric,  the real
  part of (2.4) is
\begin{equation}
\frac{\partial_t |\hat f|^2}{2}+\frac{1}{\varepsilon^2}\langle
L\hat f, \hat f\rangle = \frac{1} {\varepsilon} \,{\rm Re}\,
\langle \hat \Gamma (z,f), \hat f \rangle,
\end{equation}
 where $\langle\cdot , \cdot \rangle$
denotes the standard inner product in ${\mathbb C}^m$.

{\samepage Noting that $L$ is positive semi-definite and $\Gamma
\in N(L)^\perp$, one obtains due to (2.5) and to the
 the inequality  $ab\leq \frac{1}{2}\left(a^2 +b^2\right)$, the following
 estimate:
\begin{gather}
\frac{\partial_t | \hat f |^2}{2} +\frac{C_1}{\varepsilon^2} |
P^\perp \hat f|^2 \leq \frac{| \hat \Gamma (z,f)| |P^\perp \hat
f|}{\varepsilon} \leq \frac{1}{2C_1}|\hat \Gamma (z,f)|^2 +
\frac{C_1}{2\varepsilon^2} |P^\perp \hat f|^2,
\end{gather}
  where $C_1$ is a constant and $P^\perp$ is the orthogonal projection
onto $N(L)^\perp$.}

 In particular (2.6) implies   that
\begin{gather}
\frac{\partial_t |\hat f |^2}{2} \leq \frac{1}{2C_1}|
\hat \Gamma (z,f)|^2.
\end{gather}
Noting that for $l>\frac{n}{2}$,
\begin{gather}
\| \Gamma(f,g) \|_l \leq C_2 \| f\|_l \| g\|_l
\qquad \forall \; f,g \in H^l.
\end{gather}
So   by multiplying
(2.7) by $\left(1+|\zeta|^2\right)^l$, integrating over
$[0,t]\times {\mathbb R}_\zeta^n$ and using (2.8).  One obtains from
Plancherel's Theorem, the following inequality:
\begin{gather}
 \|  f \|^2_l \leq \|  f_0
\|^2_l +\frac{C_2^2}{C_1}\int_0^t \| f(s)\|^2_l \| z(s) \|^2_l \,ds
 \leq \|  f_0\|^2_l +C k^2 T \sup_{t\in [0,T]}\|f(t)\|^2_l,
\end{gather}
with $C=\frac{C_2^2}{C_1}$.

 Let $T_0$ and $k$ are such that
\begin{gather*}
T_0= \frac{1}{16C\| f_0\|_l^2},
\qquad k=
\frac{1-\sqrt{1-4\sqrt{CT}\|f_0\|_l}}{2\sqrt{CT}},
\end{gather*}
for $T\in[0,T_0]$.

 Then the desired estimate
(2.3) is an immediate consequence of (2.9). Thus the proof of Lemma~1 is completed.
\end{proof}

 We shall solve (1.5) by using Lemma~1 and the principle of contraction mappings. Define
the iteration scheme$\{f_\varepsilon^N\}$ by
\begin{gather}
f_\varepsilon^0=f_0,\nonumber\\
\partial_t f_\varepsilon^{N+1} + \frac{1}{\varepsilon}\sum_{j=1}^{N}
V^j\partial_{x_j}f_\varepsilon^{N+1}
+\frac{1}{\varepsilon^2}Lf_\varepsilon^{N+1}= \frac{1}
{\varepsilon}\Gamma
\left(f_\varepsilon^{N},f_\varepsilon^{N+1}\right),\nonumber\\
f_\varepsilon^{N+1}(0,x)= f_0(x), \qquad N=0,1,2\ldots
\end{gather}

\begin{lemma} Let $f_0 \in H^{l}$. Then suitable constants
 $T_0$, $k$ and $\lambda$ $(\lambda<1)$ exist
 such that
for any $\varepsilon>0$ and for any $t\in [0,T_0]$, the following
estimates are satisfied:
\begin{equation}
\left\| f_\varepsilon^{N+1}\right\|_l \leq k,
\end{equation}
 and
\begin{equation}
\left\| f_\varepsilon^{N+1}-f_\varepsilon^{N}\right\|_l \leq
C_0\lambda^{\frac{n}{2}}.
\end{equation}
\end{lemma}

\begin{proof}
Since $\left\|f^0 \right\|_l =\| f_0 \|_l
\leq k$,   (2.11) follows thanks to Lemma~1.

 Let $h_\varepsilon^{N} =  f_\varepsilon^{N+1}-f_\varepsilon^{N}$, therefore $h_\varepsilon^{N}$
satisfy the equation:
\[
\partial_t h_\varepsilon^{N} + \frac{1}{\varepsilon}\sum_{j=1}^{n}
V^j\partial_{x_j}h_\varepsilon^{N}
+\frac{1}{\varepsilon^2}Lf_\varepsilon^{N}= \frac{1}
{\varepsilon}\left(\Gamma
\left(h_\varepsilon^{N-1},f_\varepsilon^{N+1}\right) +\Gamma
\left(f_\varepsilon^{N-1},h_\varepsilon^{N}\right)\right).
\]
Applying the  technique used in  the proof of Lemma~1 one gets:
\[
\partial_t \left|\hat h_\varepsilon^N \right|^2 \leq \frac{1}{C}\left(\left| \hat \Gamma
\left(h_\varepsilon^{N-1}, \hat
f_\varepsilon^{N+1}\right)\right|^2+ \left| \hat \Gamma
\left(f_\varepsilon^{N-1},\hat
h_\varepsilon^{N}\right)\right|^2\right).
\]
Multiplying by $\left(1+|\zeta|^2\right)^l$,  integrating
 over $[0,T]\times
{\mathbb R}_\zeta^n$ and using (2.8). One can use again
Plancherel's Theorem  (2.11) to deduce that:
\[
\left\| h_\varepsilon^N\right\|_l^2 \leq C k^2 T\left(\left\|
h_\varepsilon^{N-1} \right\|_l^2 +\left\| h_\varepsilon^N
\right\|_l^2\right),
\]
with $C=\frac{C_2^2}{C_1}$.

Since $C k^2T <1$, it follows
\begin{equation}
\left\| h_\varepsilon^N \right\|_l^2 \leq
\frac{k^2cT}{1-k^2cT}\left\| h_\varepsilon^{N-1} \right\|_l^2.
\end{equation}
Put
\[
\lambda=\frac{k^2cT}{1-k^2cT}.
\]
It is clear that $\lambda<1$ and (2.12) follows from (2.13).
\end{proof}

\subsection{Proof of Theorem~1}

In view of Lemma~2 the  estimates (2.11), (2.12) imply that for
each $\varepsilon >0$, $\{f_\varepsilon^N \}$ is a~Cauchy sequence
in $L^\infty \left([0,T],H^l\right)$. Let
 denote
its limit by $f_\varepsilon (t)$ and note that it satisfies the
estimate (1.8), i.e., this limit in $L^\infty
\left([0,T],H^l\right)$.

From  (2.10)  one sees
  that  $\partial_t
  f_\varepsilon^{N+1}$ can be expressed
in terms  of sequences
converging
 in $L^\infty \left([0,T], H^{l-1}\right)$ as $N\to +\infty$.
The limit is
\[
H_\varepsilon = -\frac{1}{\varepsilon}\sum_{j=1}^{n}
V^j\partial_{x_j}f_\varepsilon
-\frac{1}{\varepsilon^2}Lf_\varepsilon+\frac{1}{\varepsilon}
\Gamma (f_\varepsilon,f_\varepsilon).
\]
Now let $\Psi (t,x)$ be a $C^\infty$
function of compact support in $[0,T]\times K$. We have just seen
that
\[
\int_{[0,T]\times K}\langle \psi (t,x), \partial_t
f_\varepsilon^{N+1} \rangle \, dt dx  \to \int_{[0,T]\times
K}\langle \psi (t,x),H_\varepsilon(t,x) \rangle \, dt dx,
\]
as $N\to +\infty$. However
\begin{gather*}
\int_{[0,T]\times K}\langle \psi (t,x), \partial_t
f_\varepsilon^{N+1} \rangle \,dt dx
 = -
\int_{[0,T]\times K}\langle \partial_t \psi (t,x),
f_\varepsilon^{N+1} \rangle \, dt dx\\ \qquad {} \to -
\int_{[0,T]\times K}\langle \partial_t \psi (t,x),  f_\varepsilon
\rangle \,dt dx,
\end{gather*}
as $N\to +\infty$. Therefore $H_\varepsilon (t)$ is identified
with the distributions derivative in $t$ of $f_\varepsilon$. It
follows that $f_\varepsilon$ satisfies the equation (1.5) and
moreover $\frac{\partial
  f_\varepsilon}{\partial   t} \in L^\infty
\left([0,T], H^{l-1}\right)$; hence $f_\varepsilon \in
C\left([0,T], H^{l-1}\right)$.
X-Mozilla-Status: 0000

\section{Hydrodynamical limit}

 This section deals with the strong
 convergence of the solution
of the discrete velocity model~(1.1) towards solution of the nonlinear
incompressible Navier--Stokes
equations (1.11)--(1.12). In order to derive the hydrodynamical limit, some uniform regularity
estimates would likely be needed for obtaining the limit of the nonlinear terms.

\subsection{Strong convergence}

 To prove Theorem~2, one needs
more than the uniform bound (1.8) for $f_\varepsilon$. For this
purpose we assume that hypothesis (1.9) holds
 and we show the uniform bound for $\frac{\partial
 f_\varepsilon}{\partial  t}$.

The uniform equicontinuity in $t$ is given by the following:

\begin{lemma}
  Assume that hypothesis (1.9) holds
  and let $l> \frac{n}{2}+1$.  Then
\begin{equation}
\left\|\frac{df_\varepsilon}{dt}\right\|_{l-1} \leq C
\exp\left(\frac{C k^2 T}{2}\right),\qquad
 \forall \; t\in[0,T],  \quad \varepsilon \in(0,1),
\end{equation}
where  the constant $C$ does not depend on $\varepsilon$.
\end{lemma}

\begin{proof} By differentiating the equation (1.5) with respect to t and by
taking the  Fourier
transform of the equation obtained, one has
\begin{equation}
\frac{d}{dt}{{\partial_t \hat f_\varepsilon}} +
\frac{1}{\varepsilon}\sum_{j=1}^{n} i \zeta_j V^j {{\partial_t
\hat f_\varepsilon}} +\frac{1}{\varepsilon^2}L {{\partial_t \hat
f_\varepsilon}}= \frac{1} {\varepsilon}\partial_t \hat \Gamma
(f_\varepsilon,f_\varepsilon).
\end{equation}
Taking the inner product (in ${\mathbb C}^m$) of (3.2) with
${\partial_t \hat f_\varepsilon}$. The real part of the resulting
equality is
\[
\frac{\partial_t |{{\partial_t \hat f_\varepsilon}} |^2}{2}
+\frac{1}{\varepsilon^2}\langle L {{\partial_t \hat
f_\varepsilon}},{{\partial_t \hat f}}\rangle = \frac{1}
{\varepsilon} \,{\rm Re}\, \langle \hat \Gamma (\partial_t f,f),
{{\partial_t \hat f}}. \rangle
\]
 Since $L$ is positive semi-definite and $\Gamma \in N(L)^\perp$, we get
\begin{gather*}
\frac{\partial_t | {{\partial_t \hat f_\varepsilon}} |^2}{2}
+\frac{C_1}{\varepsilon^2} | P^\perp {{\partial_t \hat
f_\varepsilon}} |^2 \leq \frac{|\hat\Gamma (\partial_t
f_\varepsilon ,f_\varepsilon)| |P^\perp
{{\partial_t \hat f_\varepsilon}}|}{\varepsilon}\\
\phantom{\frac{\partial_t | {{\partial_t \hat f_\varepsilon}}
|^2}{2} +\frac{C_1}{\varepsilon^2} | P^\perp {{\partial_t \hat
f_\varepsilon}} |^2 } {}\leq \frac{1}{2C_1}| \hat\Gamma
(\partial_t f_\varepsilon ,f_\varepsilon)|^2 +
\frac{C_1}{2\varepsilon^2} |P^\perp {{\partial_t \hat
f_\varepsilon}}|^2.
\end{gather*}
So,
multiplying by $\left(1+|\zeta|^2\right)^{l-1}$, integrating over
${\mathbb R}_\zeta^n$ and using (2.8) one gets
\begin{gather}
\partial_t \| \partial_t f_\varepsilon \|_{l-1}^2 \leq
\frac{1}{C_1}\| \Gamma (\partial_t f_\varepsilon ,f_\varepsilon)
\|_{l-1}^2  \leq C\| \partial_t f_\varepsilon \|_{l-1}^2 \|
f_\varepsilon \|_{l-1}^2.
\end{gather}
Using Gronwall's inequality, one
 reduces   (3.3) to
\[
\|\partial_t f_\varepsilon \|_{l-1}\leq \exp\left(\frac{C k^2
T}{2}\right) \|
\partial_t f_\varepsilon (0) \|_{l-1}.
\]
By using equation (1.5) in order to express $\partial_t
f_\varepsilon (0)$ in terms of the initial data we have
\[
 \| \partial_t f_\varepsilon \|_{l-1}\leq \exp\left(\frac{C k^2 T}{2}\right)
\left(\frac{1}{\varepsilon}\left\| \sum_{j=1}^n V^j\partial_{x_j}
f_0 \right\|_{l-1}+ \frac{1}{\varepsilon^2} \| Lf_0 \|_{l-1}
+\frac{1}{\varepsilon} \|\Gamma(f_0,f_0)\|_{l-1}\right).
\]
Taking into account hypothesis (1.9), one gets
\begin{gather*}
 \| \partial_t f_\varepsilon \|_{l-1}\leq C \exp\left(\frac{C k^2 T}{2}\right)
\left(\frac{1}{\varepsilon}\| \partial_x h_\varepsilon \|_{l-1}+
\varepsilon\| \partial_x k_\varepsilon \|_{l-1} +\|
Lk_\varepsilon\|_{l-1}\right.\\ \left. \phantom{ \| \partial_t
f_\varepsilon \|_{l-1}\leq}{}+ \varepsilon \| h_\varepsilon \|_{l}
\| k_\varepsilon \|_{l}+ \varepsilon^3  \| k_\varepsilon \|_{l}^2
+ \frac{1}{\varepsilon}
  \| \Gamma(h_\varepsilon, h_\varepsilon)\|_{l-1}\right)
  \leq C \exp\left(\frac{C k^2 T}{2}\right).
\end{gather*}
The proof of (3.1) is now complete.
\end{proof}

\noindent {\bf Conclusion.}  From Lemma 3 we conclude the
following: the solution $f_\varepsilon$ is bounded in $C([0,T],
H^{l-1})$, uniformly for $\varepsilon>0$,
 for $t$ in any compact subset of the interval $[0,T]$, moreover $f_\varepsilon$ satisfies
 the bound (3.1). Therefore by the Ascoli--Arzela lemma
 we can choose a convergent subsequence
 $f_{\varepsilon_j}$, where $\varepsilon_j \to 0$ such that
\begin{equation}
f_{\varepsilon_j} \to f \qquad \mbox{in}\quad
C\left([0,T],H^{l-1}\right).
\end{equation}
 and the limit function satisfies the bound (1.6).

\medskip

Thanks to  convergence results (3.4), the estimate (1.8)
allow us to deduce (possibly taking subsequences)
the following convergence
\begin{equation}
\Gamma(f_{\varepsilon_j}, f_{\varepsilon_j})  \to \Gamma(f,f)
\qquad \mbox{strongly in} \quad H^{l-1}.
\end{equation}

\subsection{Passage to the limit}

 This subsection is devoted to
  the connection between the discrete velocity kinetic equations~(1.1)
and the incompressible Navier--Stokes equations  (1.11)--(1.12), we use
the techniques  described in~[3]. In order to get  these
 equations, one needs
  the following:

\begin{lemma}
Let $f, g \in {\mathbb R}^m$. One gets
\begin{equation}
\langle f,Lg\rangle = \frac{1}{8} \sum_{i,j,k,l} A_{kl}^{ij}(M_i M_j +M_k M_l)
( f_i^\star+ f_j^\star -f_k^\star -f_l^\star)(g_i^\star+ g_j^\star -
g_k^\star -g_l^\star),
\end{equation}
where
\[
f_i^{\star} =\frac{f_i}{\sqrt{\alpha_i M_i}}, \qquad i=1,\ldots,m.
\]
\end{lemma}

\begin{proof} See [7] and [23].
\end{proof}

\begin{lemma} Let  $f \in N(L)$. One has
\begin{equation}
\Gamma(f,f)= \frac{1}{2}L\left(\frac{f^2}{\sqrt{\alpha M}}\right).
\end{equation}
\end{lemma}

\begin{proof} Using (1.7), one gets
\begin{gather}
\Gamma(f,f)= \Lambda^{-\frac{1}{2}}\left(\frac{1}{\alpha_i}
\sum_{j,k,l}\left( A_{kl}^{ij}\sqrt{M_k M_l}f_k f_l -
A_{ij}^{kl} \sqrt{M_i Mj}f_i f_j\right)\right)\nonumber\\
\phantom{\Gamma(f,f)}{}= \left(\sum_{j,k,l}\left( A_{kl}^{ij}\frac{\sqrt{M_k M_l}}{\sqrt{\alpha_i
\alpha_k \alpha_l M_i}}f_k f_l -
A_{ij}^{kl} \frac{\sqrt{Mj}}{\alpha_i \sqrt{\alpha_j}}f_i f_j\right)\right)_i.
\end{gather}
On the other hand, if
   $f\in N(L)$, so one gets
   by (3.6)  the following
   identity:
\[
A_{kl}^{ij}\left(\frac{f_i}{\sqrt{\alpha_i M_i}}+\frac{
f_j}{\sqrt{\alpha_j M_j}}-\frac{f_l} { \sqrt{\alpha_l
M_l}}-\frac{f_k}{\sqrt{\alpha_k M_k}}\right)=0, \qquad \forall \;
i,j,k,l.\]
 Thus, we have
 \[
 2A_{kl}^{ij}\frac{f_i
f_j}{\sqrt{\alpha_i \alpha_j M_i Mj}}=
A_{kl}^{ij}\left(\frac{f_l^2} {\alpha_l M_l}+\frac{f_k^2}{\alpha_k
M_k}-\frac{f_i^2}{\alpha_i M_i}- \frac{f_j^2}{\alpha_j
M_j}+\frac{2f_l f_k}{\sqrt{\alpha_l \alpha_k M_l Mk}}\right).
\]
Hence substituting this identity into (3.8) yields:
\begin{gather}
\Gamma(f,f)= \Bigg( \sum_{j,k,l}A_{kl}^{ij}\left(\frac{\sqrt{M_k
M_l}}{\sqrt{\alpha_i \alpha_l \alpha_k M_i}}f_k f_l-
\frac{\sqrt{M_i}M_j}{\sqrt{\alpha_i \alpha_l \alpha_k M_kM_l}}f_k
f_l-\frac{1}{2} \frac{\sqrt{M_i}M_j}{\sqrt{\alpha_i} \alpha_l M_l}
f_l^2\right.\nonumber\\ \left. \phantom{\Gamma(f,f)=}{}-
\frac{1}{2}\frac{\sqrt{M_i}M_j}{\sqrt{\alpha_i} \alpha_k M_k}
f_k^2 + \frac{1}{2}\frac{\sqrt{M_i}M_j}{\alpha_i \sqrt{\alpha_i}
M_i} f_i^2+\frac{1}{2}\frac{\sqrt{ M_i}}{\alpha_j
 \sqrt{\alpha_i}} f_j^2\right)\Bigg)_i.
 \end{gather}
Noting that \[ A_{kl}^{ij}(M_i M_j-M_k M_l)=0, \qquad \forall\;
i,j,k,l. \] So, \[ A_{kl}^{ij}\left(\frac{\sqrt{M_k
M_l}}{\sqrt{\alpha_i \alpha_l \alpha_k M_i}}f_k f_l-
\frac{\sqrt{M_i}M_j}{\sqrt{\alpha_i \alpha_l \alpha_k M_kM_l}}f_k
f_l\right) =0,
\]
  which with
 (3.9) imply
\begin{gather}
 \Gamma(f,f) = -\frac{1}{2}\left(\sum_{j,k,l} A_{kl}^{ij}
\left(\frac{\sqrt{M_i}M_j}{\sqrt{\alpha_i}\alpha_l M_l} f_l^2 +
\frac{\sqrt{M_i}M_j}{\sqrt{\alpha_i}\alpha_k M_k} f_k^2 -
\frac{\sqrt{M_i}M_j}{\alpha_i \sqrt{\alpha_i} M_i} f_i^2-
\frac{\sqrt{M_i}} {\alpha_j \sqrt{\alpha_i}}
f_j^2\right)\right)_i\! \nonumber\\ \quad{}
=-\frac{1}{2}\left(\sum_{j,k,l} A_{kl}^{ij}\left(\frac{M_k
}{\sqrt{\alpha_i M_i}} \frac{f_l^2}{ \alpha_l } +
\frac{M_l}{\sqrt{\alpha_i M_i}} \frac{f_k^2}{\alpha_k} -
\frac{M_j}{\sqrt{\alpha_i M_i}} \frac{f_i^2}{\alpha_i}-\frac{M_i}
{\sqrt{\alpha_i M_i}} \frac{f_j^2}{\alpha_j}\right)\right)_i.\!
\end{gather}
 Therefore the desired identity (3.7) follows from
(3.10) and (1.6). \end{proof}

\begin{lemma} Let $\chi$ be a test function such
that \begin{equation} \sum_{j=1}^{n}\sum_{r=1}^{d}\partial_{x_j}
\chi_r \langle e_r, V^j e_l \rangle=0, \qquad  l=1,\ldots,d.
\end{equation}
 Then
 \begin{equation}
 \langle\sum_{j=1}^{n}\partial_{x_j} \langle
f_\varepsilon, P(V^j (e_l)\rangle, \chi\rangle =0, \qquad
l=1,\ldots,d, \end{equation}
 where $P$ is the projection onto
$N(L)$.
\end{lemma}

\begin{proof} One has
\begin{gather*}
 \langle\sum_{j=1}^{n}\partial_{x_j} \langle
f_\varepsilon, P(V^j (e_l)\rangle, \chi\rangle
=\langle\sum_{j=1}^{n}\partial_{x_j} \langle Pf_\varepsilon, V^j
(e_l)\rangle, \chi\rangle\\ \qquad {} = \sum_{j=1}^{n}\langle
\partial_{x_j} \langle \sum_{r=1}^{d} \alpha_{\varepsilon, r} e_r,V^j
(e_l)\rangle,\chi\rangle  =
\sum_{j=1}^{n}\sum_{r=1}^{d}\sum_{l=1}^{d}\langle \partial_{x_j}
\alpha_{\varepsilon,r}\langle e_r,V^j (e_l)\rangle,\chi_l\rangle.
\end{gather*}
 Integrating by parts  and using (3.5), one gets
\[
 \langle\sum_{j=1}^{n}\partial_{x_j} \langle f_\varepsilon,
P(V^j (e_l)\rangle, \chi\rangle
=-\sum_{j=1}^{n}\sum_{r=1}^{d}\sum_{l=1}^{d}\langle
\alpha_{\varepsilon,r}\langle e_r,V^j
(e_l)\rangle,\partial_{x_j}\chi_l\rangle =0.
\]
This ends the proof of Lemma~6.
\end{proof}

\noindent {\bf Completion of the proof.} Multiplying the equation
(1.5) by $\varepsilon^2$, letting $\varepsilon$ go to zero and
using (3.4), yields the relation \[ Lf =0. \] This implies that
$f\in N(L)$ and thus can be written according to the formula
(1.10).

The scaled local conservation laws associated with the kinetic
equation (1.5) are
\begin{equation}
\partial_t \langle f_\varepsilon,e_i\rangle
+\frac{1}{\varepsilon}\sum_{j=1}^{n}
\partial_{x_j} \langle V^j f_\varepsilon,e_i\rangle=0,\qquad  i=1,\ldots,d.
\end{equation}
 Letting $\varepsilon$ tend to zero and applying formula (1.8)
yields condition (1.11).

As was done in [3] for the classical continuous Boltzmann
equation, we divide by $\varepsilon$ and we decompose the flux in
the form \begin{equation}
\partial_t \langle f_\varepsilon,e_i\rangle
+\frac{1}{\varepsilon}\sum_{j=1}^{n}
\partial_{x_j} \langle  f_\varepsilon,P^\perp (V^je_i)\rangle
 + \frac{1}{\varepsilon}\sum_{j=1}^{n}
\partial_{x_j} \langle  f_\varepsilon,P (V^je_i)\rangle=0.
\end{equation}

Using 3.11 (Lemma~6), the second term of (3.14) can be eliminated
upon integra\-ting~(3.14) against test functions satisfying
(3.11).

The limit of the first flux term in (3.14) is computed using the
fact that $P^\perp(V^j e_i) \in {\rm Ran}\,(L)$ along with the
kinetic equation (1.5) to obtain \begin{gather}
\partial_t \langle
f_\varepsilon,e_i\rangle +\sum_{j=1}^{n}\partial_{x_j}\langle
-\varepsilon
\partial_t f_\varepsilon - \sum_{k=1}^{n}V^k \partial_{x_k} f_\varepsilon
+\Gamma(f_\varepsilon,f_\varepsilon), L^{-1}P^\perp
(V^je_i)\rangle \nonumber\\ \phantom{\partial_t \langle
f_\varepsilon,e_i\rangle}
{}+\frac{1}{\varepsilon}\sum_{j=1}^{n}\partial_{x_j} \langle
f_\varepsilon,P (V^je_i)\rangle =0, \qquad i=1,\ldots,d.
\end{gather}
 Letting $\varepsilon $ goes to zero,
\begin{gather}
\sum_{j=1}^{n}\sum_{k=1}^{n}\partial_{x_j} \partial_{x_k}\langle
V^k f_\varepsilon,L^{-1}(P^\perp (V^je_i))\rangle \to
\sum_{j=1}^{n}\sum_{k=1}^{n}\partial_{x_j} \partial_{x_k}\langle
V^k \rho\cdot e ,L^{-1}(P^\perp (V^je_i))\rangle \nonumber\\
\qquad {} =\nabla^2 : \langle Ve \otimes
L^{-1}(P^\perp(Ve))\rangle \rho,
\end{gather}
in $D'_{t,x}$.

 The limiting quadratic term of (3.15) must be evaluated further
in order to bring it into the form that appears in (1.12). Since
$f\in N(L)$ the convergence (3.5) and the identity~(3.7) can be
employed to show the following convergence in $D'_{t,x}$ as
 $\varepsilon \to 0$,
\begin{gather}
\sum_{j=1}^{n}\partial_{x_j}\langle\Gamma(f_\varepsilon,f_\varepsilon),
L^{-1}P^\perp (V^je_i)\rangle \to
\sum_{j=1}^{n}\partial_{x_j}\langle \Gamma(f,f), L^{-1}P^\perp
(V^je_i)\rangle \nonumber\\ \qquad {}=
\frac{1}{2}\sum_{j=1}^{n}\partial_{x_j}\langle \frac{(\rho\cdot
e)^2}{\sqrt{\alpha M}},P^\perp (V^je_i)\rangle=
\frac{1}{2}\nabla\cdot \langle P^\perp (Ve),\frac{(\rho\cdot
e)^2}{\sqrt{\alpha M}} \rangle.\end{gather}

Passing to the limit $\varepsilon \to 0$ in equation (3.15), the
above convergence (3.16) and (3.17) suffices to write the limit
equation (1.12).

By integrating the equation (3.15) over $t$, we have
\begin{gather*}
 \langle f_\varepsilon,e_i\rangle - \langle f_\varepsilon
(0),e_i\rangle\\ \qquad {}=-\int_0^t\left(\sum_{j=1}^{n}
\partial x_j\langle
-\varepsilon \partial_t f_\varepsilon - \sum_{k=1}^{n}V^k
\partial_{x_k} f_\varepsilon +\Gamma(f_\varepsilon,f_\varepsilon),
L^{-1}P^\perp (V^je_i)\rangle \right.\\ \left.\qquad {}+
\frac{1}{\varepsilon}\sum_{j=1}^{n}\partial_{x_j} \langle
f_\varepsilon,P (V^je_i)\rangle(s)\right)ds=0, \qquad
i=1,\ldots,d.
\end{gather*}
Let $\varepsilon \to 0$ and putting $t=0$, it follows
\begin{equation}
 \langle f,e_i\rangle (0) =\langle
h,e_i\rangle. \end{equation}
 Therefore the desired initial
condition for the system (1.12) follows from (3.18).

In the next, we will apply the results discussed
 in Section~1 to two examples.

\section{Example, I (Carlemann model)}

The simplest one-dimensional discrete-velocity models of the Botzmann
equation are certainly those with two velocities. These models describe
the evolution of the velocity distribution of a fictitious gas composed
 of two kinds of particles
that move parallel to the $x$-axis with constant speed equal to
one, either in the positive $x$-direction with a density~$F_1$, or
in the negative $x$-direction with a density $F_2$. The
corresponding dimensionless hyperbolic system is given by (see
[20]): \begin{gather} \varepsilon{\frac{dF_1^{\varepsilon}}{dt}} +
\frac{dF_1^{\varepsilon}}{dx} =
 \frac{1}{ \varepsilon} \left( {F_2^\varepsilon}^2 - {F_1^\varepsilon}^2\right),
 \nonumber\\
\varepsilon{\frac{dF_{2}^{\varepsilon}}{dt}} -
\frac{dF_{2}^{\varepsilon}}{dx} = \frac{1}{ \varepsilon} \left(
{F_1^\varepsilon}^2 - {F_2^\varepsilon}^2\right), \nonumber\\
 F_{i}^{\varepsilon}( t=0, x ) = F_{i}^{0}, \qquad i=1.2.
 \end{gather}
  Here we have
     $(\alpha_1, \alpha_2)=(1,1)$ and
\[
A_{11}^{22}=A_{22}^{11} =1, \qquad  \mbox{and}
 \qquad A_{kl}^{ij}=0    \quad \mbox{otherwise}.
 \]
 The
 condition (1.3) is satisfied.The space $M$ of summational invariants consists
 of vectors $\phi =t_{(\phi_1,\phi_2)}$ satisfying $\phi_1 -\phi_2 =0$.
Therefore $M$ is spanned by $\phi =(1,1)$.

 On the other hand a locally Maxwellian state is a vector $
F=t_{(F_1,F_2)} >0$ satisfying $F_2^2=F_1^2$. Let $M=(1,1)$ be an
absolute Maxwellian state: $M= (1,1)$. Set $ F(t,x)=
M+\varepsilon\Lambda^{\frac{1}{2}}f(t,x)$, here $\Lambda = I_d$,
and substitute it into (4.1), we get: \begin{gather} \partial_t
f_\varepsilon + \frac{1}{\varepsilon}
\partial_{x}Vf_\varepsilon +\frac{1}{\varepsilon^2}Lf_\varepsilon= \frac{1}
{\varepsilon}\Gamma (f_\varepsilon,f_\varepsilon),\nonumber\\
f_\varepsilon(t=0,x)=f_0(x), \end{gather} where \[ L =
2\left(\begin{array}{cc} 1& -1 \\
 -1 &  1 \end{array}\right),
 \qquad  V= \left(\begin{array}{cc} 1& 0 \\
 0 &  -1 \end{array}\right),
\] and
\[
\Gamma(f,f)=\left(f_1^2 -f_2^2\right) t_{(-1,1)}.
\]

As $\varepsilon \to 0$, we get that
 $f_\varepsilon\to f$ in $D'_{t,x}$ (distribution sense)
 with  $f = \rho \cdot e$
 and $\rho$ is solution of the heat equation:
\begin{equation}
\partial_t \rho =\frac{1}{4}\partial_x^2 \rho.
\end{equation}

 It follows from the uniqueness of the solution of the initial
value problem for (4.3) that all sequences of $f_\varepsilon$ as
$\varepsilon \to 0$ give the same system (4.3) in the limit.

 Thus we have proved

\begin{theorem}  Let  $f_0\in H^1 ({\mathbb R})$, then there
exists a positive constants $T_0$ and  $k$ (depending only on $\|
f_0\|_l$) such that the initial value problem (4-2) has a unique
solution $f_\varepsilon \in L^\infty\left([0,T_0], H^1({\mathbb
R})\right)\cap C\left([0,T],L^2({\mathbb R})\right) $ satisfying
\begin{equation}
\| f_\varepsilon (t)\|_l \leq k, \end{equation}
 for $t\in
[0,T_0]$.
\end{theorem}

If in addition $f_0$ satisfies \begin{gather} f_\varepsilon (0)=
t_{( h_\varepsilon, h_\varepsilon)} +\varepsilon^2 k_\varepsilon,
\qquad \| \partial_x h_\varepsilon   \|_{2} \leq C \varepsilon,
 \qquad \lim_{\varepsilon \to 0}\| h_\varepsilon -h
\|_{2} = 0, \end{gather}
 we get

 \begin{theorem}
 As $\varepsilon \to 0$,
$f_\varepsilon\to f=\rho\cdot e $ weakly in
$L^\infty\left([0,T_0],H^1\right)$ and strongly in $ C \left( [0,
T] , L^2 (x)\right)$ and the limit satisfies the heat equation:
\begin{gather}
\partial _t \rho = \frac{1} {4}
\partial_x^2 \rho, \qquad  \rho(0,x)  = \langle h,e\rangle.
\end{gather}
\end{theorem}

\section{Example, II  (Broadwell model)}

 We shall first describe the six-velocity
 model gas considered by Broadwell [17]
 and in [29] and then specialize
 it to one dimension. We consider the following equation:
\begin{equation}
\varepsilon\frac{dF_i}{dt} + u_i\cdot \nabla F_i =
\frac{B}{3\varepsilon} ( F_{i+1} F_{i+4}+ F_{i+2}F_{i+5} -2F_i
F_{i+3} ), \qquad        i= 0,\ldots,5 \end{equation}
 Here $F_i$
are the densities of the particles with velocity \[ u_i= \left(c
\cos\left( \theta+i\frac{\pi}{3}\right), c \sin\left(
\theta+i\frac{\pi}{3}\right)\right), \]
 and $B$ denotes  the collision frequency.

We specialize (5.1) to solutions that do not depend on $y$, and
$\theta =0$, and such that $F_5= F_1$ and $F_4= F_2$ . We can
write (5.1) for $F_1$, $F_2$, $F_3$ and $F_4$:
\begin{gather}
\varepsilon\frac{dF_0}{dt} + \frac{dF_0}{dx} =
\frac{2B}{3\varepsilon} ( F_1F_2
 - F_0F_3),\nonumber\\
 \varepsilon\frac{F_1}{dt} +\frac{1}{2}\frac{dF_1}{dx} =
\frac{B}{3\varepsilon} ( F_0F_3-
 F_1F_2 ),\nonumber\\
 \varepsilon\frac{dF_2}{dt} -\frac{1}{2}\frac{dF_2}{dx} =
\frac{B}{3\varepsilon} (F_0F_3- F_1F_2  ), \nonumber\\
\varepsilon\frac{dF_3}{dt} - \frac{dF_3}{dx} =
\frac{2B}{3\varepsilon} (F_1F_2  - F_0F_3  ).
\end{gather}

 By the form of second member we have
 $(\alpha_0,\alpha_1,\alpha_2,\alpha_3)=\left(\frac{1}{2},1,1,\frac{1}{2}\right)$
  and the relation  (1,3) is satisfied.
 To get the equation (1.11)--(1.12)
 for the Broadwell's model we need some preparations.
 The space ${\mathbb M}$ of summational invariants consists
  of vectors $\phi=t_{(\phi_1,\phi_2,\phi_3,\phi_4)}$
satisfying $2\phi_1 +2\phi_4 = \phi_2+ \phi_3 $. Therefore
${\mathbb M}$ is spanned by $\{\phi_1,\phi_2,\phi_3\}$, where
\[
\phi_1= t_{(1,0,0,-1)},\qquad  \phi_2 =
t_{\left(0,1,0\frac{1}{2}\right)},\qquad \phi_3
=t_{\left(0,0,1,\frac{1}{2}\right)}\] On the other hand a locally
Maxwellian state is a vector $F=t_{(F_0,F_1,F_2,F_3)}>0$
satisfying $F_2F_1=F_0F_3$. Let $M=t_{(1,1,1,1)}$ be an absolute
Maxwellian state. Set $F(t,x)= M+ \Lambda^{\frac{1}{2}} f(t,x)$
for $\Lambda = {\rm diag}\,(2,1,1,2)$ and substitute it into
(5.2): \begin{gather}
 f_t+\frac{Vf_x}{\varepsilon} +
\frac{Lf}{\varepsilon^2}
=\frac{1}{\varepsilon}\Gamma(f,f),\nonumber\\
  f(t=0,x)=f_0(x),
  \end{gather}
where $V ={\rm diag}\left(1,\frac{1}{2}, -\frac{1}{2},-1\right)$
and \begin{gather} L = -\frac{B}{3}\left(\begin{array}{cccc} -2&
\sqrt{2}&\sqrt{2}&- 2\\
 \sqrt{2} &  -1& -1& \sqrt{2}\\
 \sqrt{2} &  -1& -1& \sqrt{2}\\
-2& \sqrt{2}&\sqrt{2}&- 2 \end{array}\right),\\
\Gamma(f,f)=\frac{B}{3}(f_2 f_1 -2f_0f_3)
t_{\left(\sqrt{2},-1,-1,\sqrt{2}\right)}. \end{gather}
 Since $N(L)
= \Lambda^\frac{1}{2} {\mathbb M}$, a simple calculations gives
the orthonormal basis $\{e_i, i=1,2,3\}$ for $N(L)$:
\[
e_1=t_{\left(\frac{\sqrt{2}}{2}, 0,0,-\frac{\sqrt{2}}{2}\right)},
\qquad e_2 = t_{\left(\frac{1}{\sqrt{10}}, \frac{2}{\sqrt{5}}, 0,
\frac{1}{\sqrt{10}}\right)}, \qquad e_3 =t_{\left(\frac{1}{\sqrt
{15}},-\frac{1}{\sqrt {30}},\frac{\sqrt{5}}{\sqrt {6}},
\frac{1}{\sqrt {15}}\right)}. \] By letting $\varepsilon \to 0$ in
(5.3) and using the results of Section~1, we get $f_\varepsilon\to
f$ in $D'_{t,x}$ (distribution sense) with $f = \rho\cdot e$,
where
\[
\rho =t_{(\rho_1 , \rho_2, \rho_3)},\qquad  e = t_{(e_1, e_2,
e_3)},
\]
 and   from (1.9) we get
\begin{equation}
\rho_3 = \frac{2\sqrt{3}}{\sqrt{10}} \rho_1, \qquad \rho_2 = -
\frac{2} {\sqrt{5}}\rho_1.
\end{equation}

\begin{lemma} Let $f=\rho \cdot e$ satisfying (5.6), then
\begin{gather}
\Gamma(f,f) =0,\\ \langle Vf,h\rangle=0 \qquad \mbox{for any}
\quad  h\in N(L). \end{gather} \end{lemma}

\begin{proof} By using (5.5) and (5.6),
we get (5.7). (5.8) is obtained by a direct calculations. This
completes the proof.
\end{proof}

\begin{lemma} Let $f= \rho\cdot e$ satisfying (5.6), then
\begin{gather}
\langle L^{-1} (P^\perp (Ve_1)), Vf\rangle  =\frac{\rho_1}{2B},\\
\langle L^{-1} (P^\perp (Ve_2)), Vf\rangle
=-\frac{\rho_1}{4B\sqrt{5}},\\ \langle L^{-1} (P^\perp (Ve_3)),
Vf\rangle  =\frac{\sqrt{6}\rho_1}{8B\sqrt{5}}.
\end{gather}
\end{lemma}

\begin{proof} The proof of this lemma follows from Lemma~7.
\end{proof}

Using the results of Section~1 combined with Lemma~8, one gets:

\begin{theorem}  Let  $f_0\in H^1 ({\mathbb R})$, then there
exists a positive constants $T_0$ and  $k$ (depending only on $\|
f_0\|_l$) such that the initial value problem (5.3) has a unique
solution $f_\varepsilon \in L^\infty\left([0,T_0], H^1({\mathbb
R})\right)\cap C\left([0,T],L^2({\mathbb R})\right) $ satisfying
\begin{equation}
\|f_\varepsilon (t)\|_l \leq k, \end{equation}
 for $t\in [0,T_0]$.
 \end{theorem}

If in addition  the initial condition satisfies:
\begin{gather}
f_\varepsilon (0)=  h_\varepsilon +\varepsilon^2 k_\varepsilon,
\qquad  Lh_\varepsilon=0, \qquad
\|\Gamma(h_\varepsilon,h_\varepsilon) \|_{2}\leq
C\varepsilon,\nonumber\\ \|\partial_x h_\varepsilon \|_{2} \leq C
\varepsilon,\qquad \lim_{\varepsilon \to 0}\| h_\varepsilon -h
\|_{2} = 0, \end{gather}
 one gets:

 \begin{theorem} As $\varepsilon \to 0$,
 $f_\varepsilon\to f=\rho\cdot e $ weakly in
$L^\infty\left([0,T_0],H^1\right)$ and strongly in $ C \left( [0,
T] , L^2 (x)\right)$ and the limit satisfies the heat equation:
\begin{gather}
\partial _t \rho_1 = \frac{1} {4B}
\partial_x^2 \rho_1, \qquad
\rho_3 = \frac{2\sqrt{3}}{\sqrt{10}} \rho_1, \qquad \rho_2 = -
\frac{2} {\sqrt{5}}\rho_1, \nonumber\\
 \rho_1(t=0) =\frac{\rho_{10}-\frac{2}
{\sqrt{5}}\rho_{20}+2\frac{\sqrt{3}}{\sqrt{10}}\rho_{30}}{3},
\end{gather}
  where
\[
\rho_{i0}= \langle h,e_i\rangle,\qquad i=1,2,3.
\]
\end{theorem}

\begin{remark} One can  remove the assumption
$\| \Gamma(h_\varepsilon,h_\varepsilon) \|_{2}\leq C\varepsilon$
in (5.13). A price to be paid for this improvement is that: we see
the solution of (5.2) in the form $F_\varepsilon = M+\varepsilon
\phi(\varepsilon) f_\varepsilon$ with $\phi(\varepsilon)
=O(\varepsilon)$ (see~[12]).
\end{remark}

\label{Bellouquid-lastpage}
\end{document}